\documentclass[12pt]{article}

\usepackage{amsmath}
\usepackage{amssymb}
\usepackage{amsfonts}
\usepackage{lscape}

\begin{document}

\fontsize{12}{6mm}\selectfont
\setlength{\baselineskip}{2em}

$~$\\[.35in]
\newcommand{\dss}{\displaystyle}
\newcommand{\raro}{\rightarrow}
\newcommand{\be}{\begin{equation}}

\def\sech{\mbox{\rm sech}}
\thispagestyle{empty}

\begin{center}
{\Large\bf Symmetry Properties of a Generalized  } \\    [2mm]
{\Large\bf Korteweg-de Vries Equation}   \\   [2mm]
{\Large\bf and Some Explicit Solutions} \\    [2mm]
\end{center}

\vspace{1cm}
\begin{center}
{\bf Paul Bracken}                        \\
{\bf Department of Mathematics,} \\
{\bf University of Texas,} \\
{\bf Edinburg, TX  }  \\
{78541-2999}
\end{center}

\vspace{3cm}
\begin{abstract}
The symmetry group method is applied to a generalized 
Korteweg-de Vries equation and several classes of group invariant 
solutions for it are obtained by means of this technique. 
Polynomial, trigonometric and elliptic function solutions
can be calculated. It is shown that  
this generalized equation can be reduced to 
a first order equation under a particular second order differential constraint
which resembles a Schr\"{o}dinger equation.
For a particular instance in which the constraint is
satisfied, the generalized equation is reduced to a quadrature. 
A condition which ensures that the reciprocal of a solution 
is also a solution is given, and a first integral
to this constraint is found.
\end{abstract}

\newpage
{\bf I. INTRODUCTION.}
\par
Recently there has been interest in the fact that 
a particular generalization of the classical 
Korteweg-de Vries (KdV)
equation can support a new type of solitary wave, which
has been referred to as a compacton in the literature.
This type of wave has a compact support and a width
which is independent of the amplitude of the wave${\bf [1,2]}$.
This equation is defined by the real parameters $(m,n)$ and 
given in terms of the function $u (x,t)$ by the fully
nonlinear KdV equation${\bf [3]}$
$$
u_{t} + ( u^m)_{x} + (u^n)_{xxx} = 0,
\eqno(1.1)
$$
The classical KdV equation has been studied extensively${\bf [4-6]}$,
in particular, by means of the inverse scattering method,
and the B\"{a}cklund transformation has been determined {\bf [7]}.
but relatively few papers have appeared which mention (1.1).

Dynamical solitons appear as a result of a balance between
weak nonlinearity and dispersion. However, it has been shown
numerically at least, that when the wave dispersion is purely
nonlinear, some novel features in the nonlinear dynamics may be
observed. The most striking and novel is the existence of a new
type of soliton, which has been referred to as a compacton in
certain quarters${\bf [3]}$. This is in contrast to the standard KdV
soliton solution, which narrows as the amplitude increases.
The width of this new type of soliton is independent of the 
amplitude.

The fully nonlinear KdV equation is of some importance
to study on physical grounds${\bf [8]}$. For example, several equations
which pertain to a discrete lattice have continuous limits
which are partial differential equations which either
resemble (1.1) or have compacton properties which are similar to
those of the generalized KdV equation. As an example,
consider a one-dimensional lattice in which each atom interacts
only with its nearest neighbors by purely anharmonic forces.
If $x_{n} (t)$ is the dimensionless displacement of the $n$ th
atom from its equilibrium position, and the atoms interact
via quartic anharmonic potentials, the equation of motion for
the $n $ th atom is given by
$$
\frac{d^2 x_{n}}{d t^2} = 
[ (x_{n+1} - x_{n})^3 +(x_{n-1} - x_{n})^3 ],
\eqno(1.2)
$$
where dimensionless units have been used. In the long and 
short-wavelength limits, the resulting partial 
differential equations have properties similar to the 
generalized KdV equation (1.1).

Not many solutions to (1.1) are known at the moment,
and so it would be useful to have whatever new results 
can be obtained at this point.
In this paper, it will be shown that the symmetry group
of equation (1.1) can be determined. By this, we mean the classical
symmetry group${\bf [9-11]}$ of equation (1.1).
In this procedure,
the coefficients of the infinitesimal generator ${\bf v}$ of
a hypothetical one parameter symmetry group are unknown
functions of $t$, $x$ and $u$. The coefficients of the
prolonged infinitesimal generator $pr^{(3)} {\bf v}$ will
be explicit expressions involving the partial derivatives 
of the coefficient functions with respect to both $t,x$ and $u$.
Eliminating any dependencies among the derivatives of $u$
modulo the original system, the coefficients of the 
remaining unconstrained partial derivatives of $u$ are equated to zero. 
This results in a large number of determining equations for 
the symmetry group which are solved. The symmetry group
will first be determined for the case $n=1$ in (1.1). 
The Lie symmetry algebra 
is found to be a subalgebra of the
usual Lie algebra for the standard KdV equation, 
which corresponds to the case $m=2$ here.
When $m=1$, the equation becomes completely linear in $u$,
and when $m=2$, the standard KdV symmetries are found.
The corresponding symmetry variables can be calculated,
and it is found that symmetry reduction to ordinary differential 
equations can be carried out. The calculation has been repeated
for the case in which $n$ is not one, and the symmetry group is
found to be much more restrictive in this case. 
The vector fields which are obtained simply correspond to 
translations in space and translations in time. 
None the less, it
is shown how some of these reductions can be used to integrate 
(1.1) and to produce 
explicit examples of solutions to (1.1).
Finally, a connection is made between (1.1) when $n=1$
and a nonlinear type of Schr\"{o}dinger equation,
which acts as a differential constraint.
This result can also be used to generate new solutions to (1.1).
It is shown how this can be done 
by obtaining a quadrature for the solution in terms
of a symmetry variable.

{\bf II. ANALYSIS OF THE EQUATION WITH ONE POWER OF u IN THE
THIRD DERIVATIVE.}

Consider the family of fully nonlinear KdV equations given by (1.1)
and written in the equivalent form with $n=1$ and $m \neq 0$ as
$$
u_{t} + u_{xxx} + m u^{m-1} u_{x} = 0.
\eqno(2.1)
$$
Three conservation laws for (2.1) exist and are 
given by
$$
u_t + ( u^m + u_{xx} )_{x} = 0,
\qquad
(u^2)_{t} + ( \frac{2m}{m+1} u^{m+1}
+ 2 u u_{xx} - u_{x}^2 )_{x} = 0,
$$
$$
( \frac{1}{m+1} u^{m+1} - \frac{1}{2} u_x^2 )_t
+ ( u^m u_{xx} + \frac{1}{2} u_{xx}^2 
 + \frac{1}{2} u^{2m} - u_x u_{xxx} - m u^{m-1} u_{x}^2 )_x =0.
$$
Of course, an equation of the form (1.1) is trivially
a conservation law itself. Let
$$
{\bf v} = \xi (x,t,u) \frac{\partial}{\partial x}
+ \tau (x,t,u) \frac{\partial}{\partial t} +
\varphi (x,t,u) \frac{\partial}{\partial u},
\eqno(2.2)
$$
be a vector field on $X \times U$. All possible coefficient functions
$\xi$, $\tau$ and $\varphi$ are to be determined
such that the corresponding one-parameter group 
$\exp (\epsilon {\bf v})$ is a symmetry group of this equation.
To obtain the prolonged equation, the operator
$$
pr^{(3)} {\bf v} =
{\bf v} + \varphi^{x} \frac{\partial}{\partial u_{x}}
+ \varphi^{t} \frac{\partial}{\partial u_{t}}
+ \varphi^{xx} \frac{\partial}{\partial u_{xx}}
+ \varphi^{xt} \frac{\partial}{\partial u_{xt}}
+ \varphi^{tt} \frac{\partial}{\partial u_{tt}}
+ \varphi^{xxx} \frac{\partial}{\partial u_{xxx}},
\eqno(2.3)
$$
is applied to equation (2.1). 
Only the terms relevant to this case
have been retained in writing (2.3). Applying (2.3) to (2.1),
it is found that the coefficient functions in 
$pr^{(3)} {\bf v}$ satisfy the equation
$$
\varphi^{t} + \varphi^{xxx} + m \varphi^{x} u^{m-1}
+ m (m-1) \varphi u^{m-2} u_{x}   =0.
\eqno(2.4)
$$
The coefficient functions in (2.3) can be calculated and
then substituted into (2.4) to  obtain
$$
\varphi_{t} - \xi_{t} u_{x} + (\varphi_{u} - \tau_{t}) u_{t}
- \xi_{u} u_{x} u_{t} - \tau_{u} u_{t}^2
$$
$$
+ D_{x}^3 \varphi - u_{x} D_{x}^3 \xi - u_{t} D_{x}^3 \tau
- 3 u_{xx} D_{x}^{2} \xi - 3 u_{xt} D_{x}^2 \tau - 3 u_{xxx} D_{x} \xi
- 3 u_{xxt} D_{x} \tau
\eqno(2.5)
$$
$$
+ m \varphi_{x} u^{m-1} + m  ( \varphi_{u} - \xi_{x}) u^{m-1} u_{x}
- m  \tau_{x} u^{m-1} u_{t} - m  \xi_{u} u^{m-1} u_{x}^2
- m \tau_{u} u^{m-1} u_{x} u_{t} + m (m-1) \varphi u^{m-2} u_{x} = 0.
$$
In (2.5), $D_{w}$ represents the total derivative of the indicated function
with respect to $w$.
Substituting the total derivatives into (2.5) and replacing the term $u_{t}$
using (2.1), the next step is to collect all like derivative terms together
and equate the coefficients to zero to obtain the set of determining equations.
The highest derivative terms $u_{xxt}$ and $u_{x} u_{xxt}$ provide the
constraints $\tau_{x} = 0$ and $\tau_{u}=0$, respectively. These
imply that $\tau= \tau (t)$ is a function of only the $t$ variable.
The coefficient of $u_{xx}^2$ gives the condition $\xi_{u}=0$, hence
$\xi$ is independent of $u$.

The terms which multiply $u_{xxx}$ give the constraint
$$
\tau_{t} = 3 \xi_{x}.
\eqno(2.6)
$$
Since $\tau$ depends only on $t$, this equation can be
integrated to obtain $\xi$ as a linear function of $x$
$$
\xi ( x, t) = \frac{1}{3} \tau_{t} (t) \, x + \sigma (t).
\eqno(2.7)
$$
The coefficients of $u_{x} u_{xx}$ and $u_{xx}$ give the
constraints $\varphi_{uu}=0$ and $\varphi_{xu}= \xi_{xx}$.
Since $\xi (x,t)$ is just linear in $x$, this pair of equations
reduces to simply $\varphi_{uu}=0$ and $\varphi_{xu}=0$.
These constraints imply that $\varphi$ is at most linear
in the function $u$, and the coefficient of $u$ a function of $t$ alone.
The remaining terms of (2.5) under these constraints is given by
$$
\varphi_{t} - \xi_{t} u_{x} - m (\varphi_{u} - \tau_{t}) u^{m-1} u_{x} 
+ m \varphi_{x} u^{m-1}  + m (\varphi_{u} - \xi_{x}) u^{m-1} u_{x}
+ m (m-1) \varphi u^{m-2} u_{x} + \varphi_{xxx} = 0.
$$
The term which multiplies $u_{x}$ is given by
$$
- \xi_{t} - m (\varphi_{u} - \tau_{t}) u^{m-1}
+ m (\varphi_{u} - \xi_{x}) u^{m-1}
+ m (m-1) u^{m-2} \varphi = 0,
\eqno(2.8)
$$
and the remaining term requires that
$$
\varphi_{t} + m u^{m-1} \varphi_{x} + \varphi_{xxx} = 0.
\eqno(2.9)
$$
Collecting like terms in (2.8), we can write
$$
- \xi_{t} + m ( \tau_{t} - \xi_{x}) u^{m-1} 
+m (m-1) \varphi u^{m-2} = 0.
\eqno(2.10)
$$
Substituting (2.7) and $\varphi (t,u) = \alpha (t) u + \beta$ into (2.10),
we obtain that
$$
- \frac{1}{3} \tau_{tt} (t) x - \sigma_{t} (t) + m (\frac{2}{3} \tau_{t}
+ (m-1) \alpha) u^{m-1} + m (m-1) \beta u^{m-2} =0.
\eqno(2.11)
$$
In order that the coefficient of $x$ vanishes, we must have
$\tau_{tt} (t) = 0$, hence $\tau (t) = c_{2} + c_{4} t$.
The coefficients of the remaining powers of $u$ must
also be equated to zero, and how this is carried out
depends on the value of $m$ to some extent.
There are three cases to consider, and we
discuss each of these cases in turn.

(i) Suppose that $m \neq 1,2$, then there are three
independent terms in (2.11), which yield the constraints
$$
\sigma_{t} (t) = 0,
\qquad
\frac{2}{3} \tau_{t} + (m-1) \alpha = 0, 
\qquad
\beta = 0.
\eqno(2.12)
$$
The general solution to (2.13) is given by
$$
\sigma = c_{1},
\quad
\alpha = - \frac{2}{3 (m-1)} c_{4},
\quad
\beta=0.
$$
Therefore, the components of the vector field can be written
$$
\xi = \frac{1}{3} c_{4} x + c_{1},
\quad
\varphi = \frac{2 c_{4}}{3 (m-1)} u,
\quad
\tau = c_{2} + c_{4} t,
$$
and the vector field (2.2) can be written
$$
{\bf v} = (\frac{1}{3} c_{4} x + c_{1}) \frac{\partial}{\partial x}
+ (c_{2} + c_{4} t) \frac{\partial}{\partial t}
- \frac{2 c_{4}}{3 (m-1)} u \frac{\partial}{\partial u}.
\eqno(2.13)
$$

(ii) When $m=0$ or $m=1$, equation (2.1) reduces to a linear equation.
When $m=1$, the last term in (2.11) is absent giving,
$$
\sigma_{t} (t) =0,
\quad
\tau_{t} (t)= 0.
$$
These imply that
$$
\sigma = c_{1}, 
\quad
\tau = c_{2},
\quad
\varphi (t,u) = \alpha (t) u + \beta.
$$
Substituting $\varphi (t,u)$ into (2.9),
it is found that $\alpha$ and $\beta$ must be
constants and the vector field is given by
$$
{\bf v} = c_{1} \frac{\partial}{\partial x} + c_{2} \frac{\partial}
{\partial t} + ( a u + b) \frac{\partial}{\partial u},
\eqno(2.14)
$$
with $a,b$ decoupled from $c_{1}$ and $c_{2}$.

When $m=0$, (2.1) is a third order linear equation
and (2.11) implies that $\tau_{tt}(t)=0$ and $\sigma_{t} (t) =0$.
Therefore, $\tau (t) = c_2 + c_4 t$ and $\sigma = c_1$.

(iii) Finally, when $m=2$, the last term in (2.11) can
be grouped with the first term to give the pair of equations
$$
\frac{2}{3} \tau_{t} (t) + \alpha =0,
\quad
\sigma_{t} (t) = 2 \beta.
\eqno(2.15)
$$
Integrating these, we find that
$$
\xi = \frac{1}{3} c_{4} x + 2 c_{3} t + c_{1},
\quad
\tau = c_{2} + c_{4} t,
\quad
\varphi = - \frac{2}{3} c_{4} u + c_{3},
$$
and the vector field (2.2) is given explicitly as follows
$$
{\bf v} = ( \frac{1}{3} c_{4} x + 2 c_{3} t + c_{1})
\frac{\partial}{\partial x}
+ ( c_{2} + c_{4} t) \frac{\partial}{\partial t}
+ ( c_{3} - \frac{2}{3} c_{4} u ) \frac{\partial}{\partial u} .
\eqno(2.16)
$$
The case $m=2$ of course corresponds exactly to the
classical nonlinear KdV equation. It can be seen then 
that for the general case $m \neq 1$, the symmetry
generators are very close in structure to the
classical KdV case. In fact, there are three 
independent generators, or vector fields specified
by (2.13), which are shown in Table 1. These
generate a Lie subalgebra of the algebra for the
classical KdV equation. Exponentiation shows that if
$u = f(x,t)$ is a solution of (2.1), then so are
the functions in the second column of Table 1.

A symmetry reduction can be carried out using ${\bf v_{3}}$
for case (i). This gives rise to the following system
$$
\frac{dx}{x} = \frac{dt}{3t} = - \frac{1}{2} (m-1) \frac{du}{u}.
$$
Integrating the first pair provides the symmetry variable $\chi$,
$$
\chi = t^{-1/3} x.
\eqno(2.17)
$$
Integrating the last pair, we obtain that
$$
u = t^{- \alpha} v (\chi),
\qquad
\alpha = \frac{2}{3 (m-1)}.
\eqno(2.18)
$$
Now $u$ in (2.18) can be differentiated with respect to
$t$ and $x$, where the derivatives of the symmetry variable
are given by $\chi_{x} = t^{-1/3}$, $\chi_{t} = - t^{-4/3} x/3$.
We obtain
$$
u_{x} = t^{-\alpha -1/3} v', \qquad u_{xxx} = t^{-\alpha -1} v''',
\qquad u_{t} = t^{-\alpha-1}
(- \alpha v - \frac{1}{3} \chi v'),
$$
where differentiation here is with respect to $\chi$.
Substituting these derivatives into (2.1), the equation
takes the form
$$
v''' + m v^{m-1} v' - \frac{1}{3} \chi v' - \alpha v = 0.
\eqno(2.19)
$$
A similar analysis can be done with ${\bf v}_{4}$ in
the third case, and it is found that the same form 
(2.19) is obtained with $\alpha=2/3$. In this case, the 
equation can be transformed into the form of a 
second Painlev\'e transcendent.

Now ${\bf v}_{1}$ and ${\bf v}_{2}$ are common in all
three cases, and this suggests that a solution of the form
$$
u(x,t) = f ( k x - \omega t),
$$
can be determined. In this case, the symmetry variable is
$y = kx - \omega t$, and writing the required
derivatives in terms of $y$, equation (2.1) can be written
$$
- \omega f_{y} + k^3 f_{yyy} + k (f^m)_{y} = 0.
\eqno(2.20)
$$
Integrating this once, it reduces to a second order equation,
$$
- \omega f + k^3 f_{yy} + k f^m = \frac{1}{2} C_{0}.
$$
Multiplying on both sides by $f_{y}$ and integrating,
we obtain the first order equation for $f$,
$$
f_{y}^{2} = C_{0} f + \frac{\omega}{k^3} f^2 - \frac{2}{k^2 (m+1)} f^{m+1}
+ \gamma.
$$
This equation can be separated and then 
integrated on both sides to give,
$$
\int \frac{df}{\sqrt{C_{0} f + \dss \frac{\omega}{k^3} f^2
- \frac{2}{ (m+1) k^2} f^{m+1} + \gamma}} = \epsilon y +a, 
\qquad 
\epsilon = \pm 1.
\eqno(2.21)
$$
For specific values of $m$, large classes of 
solutions to (2.1) can be determined from (2.21)
by varying $m$,
in particular, elliptic function solutions.

As an example, let us take $m=3$ and take $C_0 = \gamma  = 0$
in (2.21) to obtain
$$
\int \frac{df}{(f^2 ( A - B f^2))^{1/2}} = \epsilon y +a,
$$
where $A = \omega / k^3$ and $B = 1 / 2 k^2$.
Then $f$ is a solution of the following expression
$$
\frac{f ( A - B f^2)^{1/2}}{(f^2 ( A - B f^2))^{1/2} \sqrt{A}}
\ln ( \frac{2 ( A + \sqrt{A} ( A - B f^2)^{1/2})}{f} ) = \epsilon y - a.
\eqno(2.22)
$$

\newpage
{\bf III. SYMMETRIES OF THE FULLY NONLINEAR EQUATION.}

This analysis can be extended to the case of (1.1) when
$n \neq 1$. To apply $pr^{(3)} {\bf v}$ to (1.1), it
should be expanded into its constituent derivatives in
the following form,
$$
u_{t} + n (n-1) (n-2) u^{n-3} u_{x}^3 + 3n (n-1) u^{n-2} u_{x} u_{xx}
+ n u^{n-1} u_{xxx} + m u^{m-1} u_{x} = 0.
\eqno(3.1)
$$
Applying the operator $pr^{(3)} {\bf v}$ to this differential
equation, we obtain
$$
\varphi^{t} + n (n-1) (n-2) (n-3) \varphi u^{n-4} u_{x}^3 
+  3 n (n-1) (n-2) \varphi^{x} u^{n-3} u_{x}^2 + 3n (n-1) (n-2)
\varphi u^{n-3} u_{x} u_{xx}
$$
$$
+ 3 n (n-1) \varphi^{x} u^{n-2} u_{xx} + 3n (n-1) \varphi^{xx} u^{n-2} u_{x}
+n (n-1) \varphi u^{n-2} u_{xxx} + n \varphi^{xxx} u^{n-1}
\eqno(3.2)
$$
$$
+ m (m-1)  \varphi u^{m-2} u_{x} + m \varphi^{x} u^{m-1} = 0.
$$
The next step is to substitute the coefficients of the
prolongation operator into (3.2). It is understood that $u_{t}$ 
is replaced by (3.1), and the coefficients of the respective $x$
derivatives are collected and set to zero. Here, the 
expression obtained is much longer than that of the
previous example, and most of the details will be left out.
Starting with the highest derivatives $u_{xxt}$, $u_{x} u_{xxt}$
and $u_{xx}^2$, again we find that $\tau = \tau (t)$ and $\xi_{u} =0$.
The coefficient of $u_{xxx}$ is then given by
$$
u^{n-1} ( 3 \xi_{x} - \tau_{t}) - (n-1) \varphi u^{n-2} = 0.
$$
Since $n \neq 1$, the only way to eliminate the term in $u^{n-2}$
is to require that $\varphi=0$. This immediately restricts the 
form of the symmetry generator. The coefficient of $u^{n-1}$ 
must also vanish, and this gives the constraint $\tau_{t} = 3 \xi_{x}$,
which can be integrated to yield 
$\xi (x,t) = (\tau_{t}/3 ) x + \sigma (t)$.
The prolonged equation now collapses to the form
$$
- \xi_{t} u_{x}
+ n (n-1) (n-2) ( \tau_t - 3 \xi_x ) u^{n-3} u_{x}^3 
+ 3 n ( n-1) ( \tau_t - 3 \xi_x) u^{n-2} u_x u_{xx}
+ m (\tau_{t} - \xi_{x}) u^{m-1} u_{x} = 0.
$$
The second and third terms vanish due to the 
constraint $\tau_t = 3 \xi_{x}$.
This requires $\xi = c_{2}$ and $\tau = c_{1}$ 
where $c_1$ and $c_2$ are constants.
Since $\varphi=0$, the general symmetry vector field is given by
$$
{\bf v} = c_{1} \frac{\partial}{\partial t} + c_{2} \frac{\partial}{\partial x}.
\eqno(3.3)
$$
This is again a subalgebra, but even 
more restrictive than the previous case in which $n=1$.
Only the two translational symmetries survive in
this case. It is worth noting that the conclusions of this
analysis are the same for the more general
version of (1.1) given in the form
$$
u_{t} + \kappa (u^m)_{x} + \delta ( u^{n})_{xxx} = 0,
$$
where $\kappa$ and $\delta$ are real constants.

Based on the symmetry (3.3), group invariant solutions of the form
$$
u (x,t) = g( kx - \omega t),
\eqno(3.4)
$$
can be obtained. Using the symmetry variable $y= kx - \omega t$ 
and transforming the derivative into the $y$ variable,
equation (1.1) can be written in the form
$$
- \omega g_{y} + k (g^m)_{y} + k^3 (g^n)_{yyy}  = 0.
$$
Integrating once, this takes the form
$$
n ( g^{n-1} g_{y})_{y} = \frac{\omega}{k^3} g
- \frac{1}{k^2} g^{m} + C.
$$
Multiplying on both sides of this by $g^{n-1} g_{y}$,
we can integrate both sides once more to find
$$
(g^{n-1} g_{y} )^2 = \frac{2 \omega}{n (n+1) k^3} g^{n+1}
- \frac{2}{n (m+n) k^2} g^{m+n}
+ \frac{2C}{n^2} g^n + \gamma.
$$
Here, $C$ and $\gamma$ are constants of integration.
Solving this for $g_{y}$, this equation can be\
separated and then integrated to give
$$
\int
\frac{g^{n-1} \, dg}
{\sqrt{\dss \frac{2 \omega}{n (n+1) k^3} g^{n+1} 
- \dss \frac{2}{n (m+n) k^2} g^{m+n}
+ \dss \frac{2 C}{n^2} g^n + \gamma}} =
\epsilon y + a,
\quad
\epsilon = \pm 1.
\eqno(3.5)
$$
The final integration constant is written $a$,
which appears as a result of the last integration.

Let us work out the integral in (3.5) for several values of $m$ and $n$
in the case in which 
the constants of integration $C$ and $\gamma$ vanish.
It may be assumed that $\omega$ and $k$ 
are positive constants.

Consider $n=m=2$ and set $\beta= 4 \omega /3k$,
then the integral may be written in the form
$$
2k \int \frac{d g}{\sqrt{ \beta g - g^2}} =
\epsilon y + a,
\qquad
\epsilon = \pm 1.
$$
Let us take $\epsilon$ to be defined this way
in what follows.
This integral can be done, and solving for $g$,
we obtain that
$$
g( y) = \frac{\beta}{2} ( 1 + \sin (\frac{\epsilon y + a}{2 k})).
\eqno(3.6)
$$
In the case in which $a= k \pi$, using the identity
$1 + \cos 2x = 2 \cos^2 x$, this takes the form of the
compacton which has been discussed {\bf [2]}, namely,
$$
g(y) = \frac{4 \omega}{3 k} \cos^2 (\frac{y}{4 k}).
\eqno(3.7)
$$
This solution has the property that it is positive
for $| y | <2 \pi k$ and zero at the endpoints.
This fact enables us to define a compacton form
of solution by taking a solution of the form
$$
\begin{array}{ccc}
   &  \displaystyle \frac{4 \omega}{ 3 k} \cos^2 
( \displaystyle \frac{kx -  ct}{4k} ), & \quad |kx - ct | < 2 \pi k,  \\
u (x,t) =  \Large\{ &     &  \\
   &     0,     &  \quad |kx - ct | > 2 \pi k.   \\
\end{array}
$$
Moreover, the derivative of this $u (x,t)$ has
a derivative that is continuous at the endpoints of this interval.

Let $n=3, m=2$, then the integral reduces to
$$
\sqrt{\frac{15}{2}} k \int \frac{dg}{\sqrt{\dss \frac{5 \omega}{4k} -g}}
= \epsilon y + a.
$$
It follows that $g(y)$ is given by
$$
g(y) = \frac{5 \omega}{4 k} - \dss \frac{1}{30 k^2} (y+a)^2.
\eqno(3.8)
$$

Let $n=2, m=3$, then with $\beta= 5 \omega/ 3k$, the
integral takes the form,
$$
\sqrt{\frac{5}{2}} k \int \frac{dg} 
{\sqrt{\beta g - g^3}} = \epsilon y + a.
$$
Integrating and solving for $g$, a Jacobi elliptic
function is obtained as a solution to (1.1) in this case,
$$
g (y) = \beta ( sn ( \frac{\sqrt{-5 \sqrt{\beta}}(\epsilon y + a)}{2 k},
\frac{1}{\sqrt{2}})^2 -1).
\eqno(3.9)
$$

Finally, for $n=3$, $m=3$ we set $\beta= 3 \omega / 2 k$, then the
integral takes the form
$$
3 k \int \frac{dg}{\sqrt{ \beta - g^2}} =
\epsilon y + a.
$$
Integrating and solving for $g(y)$, we have
$$
g (y) = \sqrt{\beta} \sin ( \frac{\epsilon y +a}{3 k}).
\eqno(3.10)
$$
Other cases could be integrated and would provide
elliptic function solutions to (1.1).

{\bf IV. REDUCTION OF EQUATION SUBJECT TO A DIFFERENTIAL CONSTRAINT.}

An interesting reduction of (2.1) takes place if we
subject it to a differential constraint which has a structure
analogous to that of a Schr\"{o}dinger
equation. To generate further solutions of (2.1), the
following proposition can be used {\bf [12]}.

{\bf Proposition 1.} Let $f(x,t)$ and $g(x,t)$ be functions which
satisfy the differential equation
$$
\frac{\partial^2 \psi}{\partial x^2} + [ \frac{m}{2 (m+1)} u^{m-1}
+ \lambda ] \psi = 0,
\eqno(4.1)
$$
where $u$ is defined as 
$$
u(x,t) = f(x,t) \cdot g(x,t),
\eqno(4.2)
$$
$\lambda$ is a real constant. Then
the generalized KdV equation (2.1) reduces to the form of
the first order partial differential equation given as follows
$$
f ( \frac{\partial g}{\partial t} - 4 \lambda \frac{\partial g}
{\partial x} ) + g ( \frac{\partial f}{\partial t} - 4 \lambda \frac{\partial f}{\partial x}) = 0.
\eqno(4.3)
$$

{\bf Proof.} With $u$ defined by (4.2), let us write the last 
nonlinear term in (2.1) in the form
$$
 m u^{m-1} u_{x} =  b ( f g)^{m-1} \frac{\partial u}{\partial x}
+ a u^{m-1}
( f \frac{\partial g}{\partial x} + g \frac{\partial f}{\partial x}),
$$
where $a$ and $b$ are constants which satisfy $a+b=m$.
Differentiating $u$ with respect to $x$ and $t$,
equation (2.1) takes the following form
$$
f \{ \frac{\partial g}{\partial t} +
\frac{\partial^3 g}{\partial x^3} + \frac{3}{f}
\frac{\partial^2 f}{\partial x^2} \frac{\partial g}{\partial x}
+ a u^{m-1} \frac{\partial g}{\partial x}
+ \frac{b}{2} f^{m-2} g^{m-1} \frac{\partial u}{\partial x} \}
$$
$$
+ g \{ \frac{\partial f}{\partial t} +
\frac{\partial^3 f}{\partial x^3} + \frac{3}{g}
\frac{\partial^2 g}{\partial x^2} 
\frac{\partial f}{\partial x}
+ a u^{m-1} \frac{\partial f}{\partial x} + \frac{b}{2}
f^{m-1} g^{m-2} \frac{\partial u}{\partial x} \} = 0.
\eqno(4.4)
$$
Suppose $f$ and $g$ are required to satisfy the equation
$$
\frac{\partial^2 \psi}{\partial x^2} 
- [ q u^s - \lambda] \psi = 0,
$$
where $u$ is given by (4.2). Let us show that
we can pick $q$ and $s$ in a unique way such that the
conclusion of the proposition holds.
The third derivatives of $f$ and $g$ can be obtained 
by differentiating this constraint
with respect to $x$. Substituting 
these derivatives into (4.4), the quantity inside the first bracket 
in (4.4) can be written in the form
$$
\frac{\partial g}{\partial t} + (q s f^{s-1} g^s + \frac{b}{2}
f^{m-2} g^{m-1}) \frac{\partial u}{\partial x}
+ ( 4 q u^s  + a u^{m-1} ) \frac{\partial g}{\partial x}
- 4 \lambda \frac{\partial g}{\partial x}.
$$
If we take $s= m-1$, then the coefficients of the second and
third terms reduce to the system of equations
$$
q (m-1) + \frac{b}{2} = 0,
\quad
4 q + a =0.
$$
Solving these equations subject to the condition that
$a+b=m$, we obtain the solution
$$
a= \frac{2m}{m+1},
\quad
b = \frac{m (m-1)}{m+1},
\quad
q = - \frac{m}{2 (m+1)}.
\eqno(4.5)
$$
This procedure can be repeated on the second bracket in
(4.4) and exactly the same solution (4.5) for these constants 
is obtained. Therefore, combining the remaining terms in (4.4) 
then gives the result (4.3). QED.

Now dividing both sides of (4.3) by (4.2),
it is easy to see that, using the linearity of the 
derivative operators, (4.3) can be put in the
equivalent form
$$
( \frac{\partial}{\partial t} - 4 \lambda \frac{\partial}{\partial x} )
\ln ( f(x,t) \cdot g(x,t) ) = 0.
$$
This result implies that $u (x,t)$ has the particular structure
$$
u (x,t) = f(x,t) \cdot g(x,t) = h ( x + 4 \lambda t),
$$
where the function $h$ is unspecified for the moment.

Let us consider an instance in which $h$
can be determined explicitly by 
making use of the constraint (4.1) as an example. 
Consider the case in which
$$
f(x,t) = g(x,t)= \phi (x,t).
$$ 
From the preceding considerations,
$\phi (x,t)$ must have the form
$\phi (x,t) = \phi ( x + 4 \lambda t) = \phi ( y)$,
where $y = x + 4 \lambda t$. Of course, $\phi (x,t)$ must
satisfy the second order equation (4.1) as well, which takes the form,
$$
\ddot{\phi} + \frac{m}{2 (m+1)} \phi^{2m-1} + \lambda \phi = 0.
\eqno(4.6)
$$
Differentiation in (4.6) is with respect to the symmetry variable $y$.
The constraint is sufficient to determine the function $h$ in this case.
Multiplying (4.6) by $\dot{\phi}$, this can be integrated to give
$$
\dot{\phi}^{2} = C_{0} - \frac{1}{2 (m+1)} \phi^{2m}
- \lambda \phi^2,
$$
where $C_{0}$ is a constant of integration. This equation can
finally be integrated by quadrature to give the integral
$$
\int \frac{d \phi}{\sqrt{C_{0} - \dss \frac{1}{2 (m+1)} \phi^{2m} 
- \lambda \phi^2}} = y + a.
\eqno(4.7)
$$
The integral (4.7) will generate solutions to (2.1)
in the form of elliptic functions. As an example,
(4.7) can be integrated when $m=1$ to give a solution
$$
\phi(x,t) = 
2 \sqrt{\frac{C_{0}}{4 \lambda +1}} 
\sin (\frac{1}{2} \sqrt{4 \lambda +1} 
( x + 4 \lambda t +a)).
\eqno(4.8)
$$
Squaring (4.8), we obtain an explicit solution to (2.1) of the 
form $u (x,t) = \phi (x,t)^2$.

Let us now generalize a result in {\bf [12]} to the case of (2.1).

{\bf Proposition 2.} Suppose that $w = w(x,t)$ is a
solution to the generalized KdV equation (2.1) and 
satisfies the additional constraint
$$
\frac{\partial^2 w}{\partial x^2} - \frac{1}{w} ( \frac{\partial w}
{\partial x} )^2 + \frac{m}{6} w ( w^{m-1} - w^{-m+1}) =0.
\eqno(4.9)
$$
Then the reciprocal $u=1/w$ is a solution to (2.1)

To prove this, substitute $ u (x,t) = 1 / w(x,t)$ into (2.1).
Replacing $w_t$ from (2.1) and $w_{xx}$ from (4.9), 
the result follows.

A first integral for (4.9) can be obtained of the form
$$
( \frac{\partial w}{\partial x})^2 = a w^{m+1} + b w^{-m +3} + K w^2,
\eqno(4.10)
$$
where $a$ and $b$ will depend on $m$ and $K$ is a constant
of integration. Differentiating both sides with respect to
$x$, we obtain
$$
\frac{\partial^2 w}{\partial x^2} = \frac{a}{2} (m+1) w^m - \frac{b}{2} (m-3) w^{-m+2} + K w.
\eqno(4.11)
$$
An expression for $K w$ can be obtained from (4.10),
and substituting this into (4.11), there results
$$
\frac{\partial^2 w}{\partial x^2} = 
\frac{1}{w}( \frac{\partial w}{\partial x})^2 + (\frac{m-1}{2}) a w^m - (\frac{m-1}{2}) b w^{-m+2}.
$$
Comparing this to (4.9), it must be that $a =b = -m/ (3 (m-1))$ when $m \neq 1$. 
Thus, we have proved the following claim.

{\bf Proposition 3.} When $m \neq 1$, the equation
$$
( \frac{\partial w}{\partial x})^2 = - \frac{m}{3 (m-1)} w^2 ( w^{m-1}
+ w^{-m+1})  + K w^2,
\eqno(4.12)
$$
is a first integral for (4.9).

More can be said with regard to the class of functions
referred to in Proposition 2. This will generalize what 
was done in {\bf [12]} for the usual KdV equation.
Differentiating the
first integral in (4.11) with respect to $x$,
another relation for $w_{xxx}$ results
$$
w_{xxx} = \frac{a}{2} m ( m+1) w^{m-1} w_{x}
+ \frac{b}{2} (m-3) (m-2) w^{-m+1} w_{x} + K w_{x}.
\eqno(4.13)
$$
Replacing the third derivative (4.12) in equation
(2.1), a first-order nonlinear equation is obtained
$$
( \alpha w^{m-1} + \beta w^{-m+1} + K) w_x + w_t = 0,
\eqno(4.14)
$$
where
$$
\alpha = \frac{a}{2} m (m+1) +m,
\qquad
\beta = \frac{b}{2} ( m-3) (m-2).
$$
Equation (4.14) is a quasilinear equation and the
following initial value problem can be solved
$$
\tau ( w ) w_{x} + w_{t} = 0,
\qquad
w ( x, 0) = w_{0} (x),
$$
where $\tau ( w ) = \alpha w^{m-1} + \beta w^{-m+1} +K$.
Consider the equivalent problem
$$
\begin{array}{ccc}
x_{r} = \tau (w), &  t_r = 1,  &   w_{r} = 0  \\
   &     &   \\
x (0,s) = s,  &  t (0, s) = 0,  &  w( 0, s) = w_{0} (s).
\end{array}
\eqno(4.15)
$$
Here $w_{0}$ is an arbitrary function of one-variable
for the moment. Integrating this first order system,
we obtain the result $w = w_{0} (s)$, $t=r$ and $s = x - 
\tau (w) t$, from which it follows that
$$
w = w_{0} ( x - \tau (w) t).
\eqno(4.16)
$$
This will actually furnish the solution to (4.15)
provided that the equation $\Phi (x,t,w )
= w - w_{0} ( x - \tau (w) t)=0$ can be solved for
$w$ as a function of $x$ and $t$. Substituting (4.16)
into (4.10), it is reduced to an ordinary differential equation
$$
\dot{w}_{0}^2 = a w_{0}^{m+1} + b w_{0}^{-m+3} + K w_{0}^2,
$$
which can be put in the form of a quadrature
$$
\int \frac{d w_0}{( a w_{0}^{m+1} + b w_{0}^{-m+3} + K w_{0}^2)^{1/2}}
= \pm ( x - \tau (w) t) +c.
\eqno(4.17)
$$
This will determine the class of functions $w_0$ which
will determine $w$ by solving (4.16) and satisfy
Proposition 2.

The analogue of Proposition 2 for equation (1.1) is as follows.

{\bf Proposition 4.} Suppose that $w = w (x,t)$ is a solution
of the generalized KdV equation (1.1) and in addition, 
satisfies the third order constraint
$$
( w^n - w^{-n+2}) \frac{\partial^3 w}{\partial x^3}
+ [ 3 ((n+1) w^{-n+1} + (n-1) w^{n-1})
\frac{\partial^2 w}{\partial x^2} + \frac{m}{n} 
( w^m - w^{-m+2}) ]
\frac{\partial w}{\partial x}
$$
$$
+ ( w^{n-2} ( n^2 - 3n +2) - w^{-n} (n^2 + 3n +2))
(\frac{\partial w}{\partial x})^3 =0.
\eqno(4.18)
$$
Then the reciprocal function $u=1/w$ is also a
solution of (1.1). When $n=1$, equation (4.18) reduces to (4.9).

{\bf Proposition 5} When $m=n$ in equation (1.1),
there exists a separation of variables solution of the form
$$
u (x,t) = f(x) \cdot g(t),
\eqno(4.19)
$$
provided that $f$ and $g$ can be found which satisfy the
equation
$$
\begin{array}{c}
(f^n)_x + ( f^n)_{xxx} + \lambda f = 0,  \\
  \\
g_t - \lambda g^n = 0.  \\
\end{array}
\eqno(4.20)
$$
The second equation in (4.20) can be integrated to give $g (t)$ 
from $ (n-1) g^{n-1} = - ( \lambda t + c)^{-1}$ when $n \neq 1$
and $g(t) = c e^{\lambda t}$ when $n=1$ and $c$ is a constant here.

{\bf V. SUMMARY.}

To conclude, the symmetry group for the generalized KdV 
equation has been calculated. The translational symmetry
which was found, although of frequent occurrence in such types
of equations, for the case of compacton solutions leads
to the idea of using this system to model sets of
bubbles and droplets or bubble patterns.
Thus, one can imagine sequences of bubbles which are juxtaposed in some order,
such that the pattern can be translated into itself.
In classical soliton theory,
integrability and elastic collisions are closely connected.
Some conservation laws have been found for (1.1)
previously, but it is not known whether the equation is
integrable ${\bf [2]}$. It might be hoped that symmetry methods
can be useful in searching for new conservation laws,
and perhaps to help settle the question of integrability 
for this system.

{\bf Acknowledgement}

I would like to thank Professor L. Debnath for suggesting the
study of this equation to me.

\newpage
\vspace{1cm}
{\bf References.}

\noindent
$[1]$ A. J. Sievers and S. Takeno, Intrinsic Localized Modes 
in Anharmonic Crystals, Phys. Rev. Letts., {\bf 61}, 970-973,
(1998).   \\
$[2]$ P. Rosenau and J. Hyman, Compactons: Solitons with
Finite Wavelength, Phys. Rev. Letts., {\bf 70}, 564-567, (1993). \\
$[3]$ Y. S. Kivshar, Intrinsic localized modes as solitons
with a compact support, Phys. Rev {E 48}, R43-45, (1993).   \\
$[4]$ A. Das, Integrable Models, World Scientific Notes
in Physics, Vol. 30, (1989). \\
$[5]$ M. J. Ablowitz and H. Segur, Solitons and the Inverse
Transform Method, (SIAM Philadelphia, 1981).   \\
$[6]$ R. M. Miura, The Korteweg-de Vries Equation:
A Survey of Results, Siam Review, {\bf 18}, 412-459 (1976).  \\
$[7]$ A. C. Newell, The Interrelation between B\"{a}cklund
Transformations and the Inverse Scattering Transform, in Lecture Notes
in Mathematics, vol. 515, edited by R. M. Muira, Springer-Verlag,
Berlin, 1976.  \\
$[8]$ A. Ludu and J. P. Draayer, Patterns on liquid surfaces:
cnoidal waves, compactons and scaling, 
Physica {\bf D 123}, 82-91, (1998).  \\
$[9]$ P. J. Olver, Applications of Lie Groups to Differential 
Equations, 2nd Ed. (Springer Verlag, New York, 1993).  \\
$[10]$ P. J. Olver, Symmetry and explicit solutions of
partial differential equations, Appl. Numerical Math.
{\bf 10}, 307-324, (1992).   \\
$[11]$ P. J. Olver and P. Rosenau, Group-invariant solutions 
of differential equations, SIAM J. Appl. Math. {\bf 47},
263-278, (1987).  \\
$[12]$ P. Bracken, Some methods for generating solutions to
the Korteweg-de Vries equation, Physica {\bf A 335}, 70-78, (2004). \\

\newpage

{\bf Table 1.} Symmetry algebra spanning vector fields
and exponentiated solutions for equation (2.1). The case $m=2$
corresponds to the usual KdV equation.

\vspace{3mm}
\begin{tabular}{|l|r|r|}    \hline
Case   & Symmetry Vector Field  & Exponentiated Solution    \\   \hline
       &    &       \\
$m \neq 1,2$     & ${\bf v}_{1} = \dss \frac{\partial}{\partial x}$   &
$u^{(1)} = f(x- \epsilon, t)$     \\ [2mm]
       & ${\bf v}_{2} = \dss \frac{\partial}{\partial t}$   &
$u^{(2)} = f(x, t- \epsilon)$     \\  [2mm]
       & ${\bf v}_{3} = x \dss \frac{\partial}{\partial x}
+ 3 t \dss \frac{\partial}{\partial t} - \dss \frac{2}{m-1} u 
\frac{\partial}{\partial u}$ & $u^{(3)} = 
f ( e^{- \epsilon} x, e^{-3 \epsilon} t) e^{-2 \epsilon/(m-1)}$       \\ [2mm]
       &    &      \\
       &    &      \\   \hline
       &    &      \\
$m =1$     & ${\bf v}_{1} = \dss \frac{\partial}{\partial x}$   &
$u^{(1)} = f (x - \epsilon,t)$    \\    [2mm]
       & ${\bf v}_{2} = \dss \frac{\partial}{\partial t}$   &
$u^{(2)} = f(x, t- \epsilon)$     \\    [2mm]
       &  ${\bf v}_{3} = (u +1) \dss \frac{\partial}{\partial u}$ &
$u^{(3)} = e^{-\epsilon} f (x,t) + \epsilon$   \\   [2mm]
       &   &       \\
       &   &       \\  \hline
       &   &        \\
$m=2$     & ${\bf v}_{1} = \dss \frac{\partial}{\partial x}$   &
$u^{(1)} = f(x- \epsilon , t)$   \\   [2mm]
       & ${\bf v}_{2} = \dss \frac{\partial}{\partial t}$   &
$u^{(2)} = f (x, t- \epsilon)$  \\   [2mm]
       & ${\bf v}_{3} = 2 t \dss \frac{\partial}{\partial x} +
\dss \frac{\partial}{\partial u}$  & $u^{(3)} = 
f (x - 2 \epsilon t,t) + \epsilon$       \\   [2mm]
       & ${\bf v}_{4} = x \dss \frac{\partial}{\partial x} 
+ 3 t \dss \frac{\partial}{\partial t} - 2 u \dss \frac{\partial}{\partial u}$  &
$u^{(4)} = e^{-2 \epsilon} f ( e^{- \epsilon} x, e^{-3 \epsilon} t)$  \\  
[2mm]  \hline
\end{tabular}
\end{document}